\documentclass{PoS}

\usepackage[square,comma,numbers,sort&compress]{natbib} 
\usepackage[utf8]{inputenc}
\usepackage{graphicx}
\usepackage{color}
\usepackage{amsfonts}
\usepackage{amsmath}

\title{{\vspace{-18mm} \normalsize\hfill{\small DESY 15-205}}\\[10mm]
Lattice study of the Higgs-Yukawa model in and beyond the Standard Model}

\ShortTitle{Lattice study of the Higgs-Yukawa model in and beyond the Standard Model}

\author{\speaker{David Y.-J. Chu} 
         \\
        National Chiao-Tung University, Hsinchu, Taiwan\\
        E-mail: \email{ren1072.ep99@g2.nctu.edu.tw}}

\author{Karl Jansen\\
       NIC, DESY Zeuthen, Germany\\
       E-mail: \email{karl.jansen@desy.de}}

\author{Bastian Knippschild\\
       HISKP, Bonn, Germany\\
       E-mail: \email{knippschild@hiskp.uni-bonn.de}}

\author{C.-J. David Lin\\
       National Chiao-Tung University, Hsinchu, Taiwan\\
       E-mail: \email{dlin@mail.nctu.edu.tw}}

\author{Kei-Ichi Nagai\\
       Kobayashi-Maskawa Institute, Nagoya University, Japan\\
       E-mail: \email{keiichi.nagai@kmi.nagoya-u.ac.jp}}
       
\author{Attila Nagy\\
       Humboldt University Berlin; NIC, DESY Zeuthen, Germany\\
       E-mail: \email{nagy@physik.hu-berlin.de}}

\abstract{
We derive finite-size scaling formulae for four-dimensional
Higgs-Yukawa models near the Gaussian fixed point.  These formulae
will play an essential role in future, detailed investigation of such
models.  In particular, they can be used to determine the nature of
the observed phase transitions, and confirm or rule out the
possibility of having non-trivial fixed points in the Higgs-Yukawa
models.  Our scaling formula for Binder's cumulant is tested against
lattice simulations carried out at weak couplings, and good agreement
is found.   As a separate project, we also present preliminary results
from our study of a chirally-invariant Higgs-Yukawa model including a
dimension-six operator at finite temperature.   Our work provides first
indications of first-order temperature-induced phase transitions near 
the infinite cutoff limit in this model.
}

\FullConference{The 33rd International Symposium on Lattice Field Theory\\
                 14 -18 July  2015\\
                 Kobe International Conference Center, Kobe, Japan}

\begin{document}

\section{Introduction}
The Standard Model (SM) has been a successful theory for explaining interactions amongst elementary particles.  However the 
scalar sector of the SM is trivial and hence the cutoff cannot be removed, leading to the need of new physics beyond 
the cutoff scale.  The situation may be different as we investigate the Higgs-Yukawa (HY) sector.  Recent works 
using perturbation theory \cite{Molgaard:2014mqa,Hung:2009hy}, as well as lattice studies \cite{Bulava:2012rb},
suggest the HY models may contain non-trivial 
fixed points that lead to UV completion. 
We would like to explore this scenario by performing lattice simulations.
The main task is to be able to distinguish between the Gaussian and the strongly-coupled fixed points.
In order to achieve this, we rely on the method of
finite-size scaling (FSS).

Finite-size scaling provides a link between the critical exponents of a phase transition in infinite volume theory 
and the finite-volume scaling behaviour.
This method allows us to identify different universality classes from extracting the 
critical exponents by studying the theory near second-order phase transitions. 
It is well known that in the 4-dimensional pure scalar theory, scaling behaviour near the Gaussian fixed 
point (GFP) receives logarithmic corrections \cite{Frohlich:1982tw}.
In our current work, we determine the leading logarithmic corrections by performing one-loop calculations,
and test the result for
Binder's cumulant against lattice simulations for a chirally-invariant HY model
with two mass degenerate fermions.

As a separate project, we also present preliminary results for finite temperature properties of the HY model with a 
dimension-6 term, $(\Phi^{\dagger}\Phi)^3$. 
At low cutoff around a few TeV, effects of this operator on low energy observables may not be negligible 
\cite{Chu:2015nha,Akerlund:2015fya,Eichhorn:2015kea,Branchina:2013jra}.
Such a term is in principle allowed in the presence of cutoff, and can be understood as a proxy for an extension of the SM.
We study the finite temperature property and find indications for 
temperature-induced first-order phase transitions near the infinite cutoff limit.  
If such first-order phase transitions persists in the full SM and turns out to be strong 
enough, they can lead to Baryogensis \cite{Cohen:1993nk}. 
\section{Finite Size Scaling near the Gaussian Fixed Point in the Higgs-Yukawa Model}
Universality classes of a theory can be characterised by the 
renormalisation group (RG) running behaviour of the couplings near critical points.  
The GFP in four dimension is special as the beta function contains a double zero, 
leading to logarithmic scaling behaviour \cite{Frohlich:1982tw}.
Below we study such scaling behaviour for the HY models.
In our calculations, anisotropic lattices have been used with the volume being $L^3 \times L_t$, 
where $L_t=2L$ is the Euclidean time direction.  
In this work we investigate the HY model with scalar quartic coupling $\lambda$ and degenerate Yukawa interaction $y$.
The scalar mass square is denoted as $m^2$.  We introduce $Y=y^2$. 
This coupling, $Y$, has the same classical dimension as $\lambda$ in arbitrary space-time dimension.
Below we denote the bare quantities with subscript $b$.  The quantities with a hat
are measured in units of the lattice spacing $a$.

Consider a bare correlator, $M_b(m_b^2 , \lambda_b, Y_b; a, L)$, at zero momentum, 
with  classical dimension $D_M$. It depends on the bare couplings, $m_b^2$, $\lambda_b$, $Y_b$, the
lattice spacing, $a$, and the box length, $L$.  The RG analysis leads to
\begin{eqnarray}
 \hat{M}_b \left[ m_b^2 , \lambda_b, Y_b; a, L \right] Z_{\phi}^{-D_{M}/2} (a,l) 
 &=& \hat{M} \left[ m^2 (\hat{l}), \lambda (\hat{l}), Y(\hat{l}); l, L \right],  \nonumber \\
\label{renormalised_rescaled_quantity}
 &=& \zeta_{M} (l,L) \hat{L}^{-D_{M}} \hat{M} \left[ \hat{m}^2 (\hat{L}) \hat{L}^{2}, \lambda (\hat{L}), Y(\hat{L});1,1 \right], \\
\text{where } \zeta_M (l,L) &=& \exp \left( \int_{l}^{L} \gamma_M (\rho) \mathrm{d} \log \rho \right),
\end{eqnarray}
$l$ is a length scale that is assumed to be significantly larger than the lattice spacing.
In Eq.~(\ref{renormalised_rescaled_quantity}), the first step is the implementation of (non-perturbative)
matching from the bare correlator to its renormalised counterpart at the renormalisation scale $l$.
The second step is the RG running from $l$ to $L$ through $\zeta_M$, 
where $\gamma_{M}$ is the anomalous dimension of the correlator $M$. 
Near a fixed point, the marginal couplings, $\lambda (\hat{L}) \text{ and } Y(\hat{L})$, in
Eq.~(\ref{renormalised_rescaled_quantity}) approach constants.  
Aside from the prefactors, the correlator in Eq.~(\ref{renormalised_rescaled_quantity}) 
only depends on the dimensionless combination $\hat{m}^2(\hat{L})\hat{L}^2$.
The RG analysis in Eq.~(\ref{renormalised_rescaled_quantity}) does not give us the detailed functional form 
of $\hat{M} \left[ \hat{m}^2(\hat{L}) \hat{L}^2 \right]$.
Below we perform analysis to obtain this functional form
for the HY model near the GFP.

Consider the partition function of the HY model which contains scalar $O(N)$ and fermion $SU(N_f)$ flavour symmetries 
with scalar quartic and degenerate Yukawa interactions.
Using the notation $\Phi^T = (\phi_1 , \dots, \phi_N)$ and $\Psi^T = (\psi_1, \dots , \psi_{N_f})$,
where the transpose is in flavour space, the partition function of this theory is,

\begin{equation}
\label{full_partition_function}
 Z=\int D\Phi \mbox{ } D\bar{\Psi} \mbox{ } D\Psi \exp \left( -S[\Phi, \bar{\Psi}, \Psi] \right).
\end{equation}
Since we are interested in finite volume effects, after performing the fermionic integrals,
we separate the scalar fields into
$\phi_a = \varphi_a + \chi_a$, where $\varphi_a$ are the zero modes.  Near the GFP,
$\chi_a$ can be treated perturbatively, and contribute through loop effects.  
This leads to,
\begin{eqnarray}
\label{partition_function_after_zero_mode_seperation}
 Z=\int_{-\infty}^{\infty} \mathrm{d}^N \varphi_a \, \mathcal{N} \exp (-S_{eff}[\varphi_{a}])
 =\Omega_{N-1} \int_{0}^{\infty} \mathrm{d} \varphi \, \varphi^{N-1}  \mathcal{N} \exp (-S_{eff}[\varphi]),
\end{eqnarray}
where $\mathcal{N}$ is the contribution from the non-Gaussian modes of $\chi_a$, 
and $S_{eff}$ is the effective action containing the result 
of the Gaussian integrals of the $\chi_a$ and the fermions.
In Eq.~(\ref{partition_function_after_zero_mode_seperation}), we have written 
$\varphi^2 = \sum_{a=1}^N \varphi_a^2$, and the $N{-}$dimensional integral of the zero modes
is turned into a one-dimensional integral with the solid angle denoted as $\Omega_{N-1}$.  
In this work, we are only studying the theory to one-loop order, 
where $\mathcal{N}$ does not result in the renormalisation
of the couplings and the fields in $S_{eff}[\varphi]$, and can be regarded as an overall constant.
The effective action only depends on $\varphi$ because of the $O(N)$ symmetry.
In the vicinity of the GFP, $S_{eff}$ can be studied using the saddle point expansion around the zero mode \cite{Brezin:1985xx}.
We start discussing this expansion by writing,
\begin{equation}
\label{effective_action_first_expansion_to_mean_field}
 \exp ( -S_{eff}[\varphi] ) =  \mathrm{det} \left( M_F [\varphi] \right) 
 \mathrm{det} \left( M_B [\varphi] \right)^{-1}  \exp \left( -L^4m_b^2 \varphi^2 -2L^4\lambda_b \varphi^4 \right),
\end{equation}
where the determinants of the matrices $M_B[\varphi] \text{ and } M_F[\varphi]$ 
come from the Gaussian integrals of the scalar non-zero modes and the fermions, respectively.
Expanding the determinants results in the renormalisation of the couplings, 
$m^2(\hat{L}), \, \lambda(\hat{L})$, and $Y(\hat{L})$.  
It also leads to the volume-dependent additive renormalisation of $m^2(\hat{L})$, as 
well as the appearance of higher dimensional operators composed of the zero mode, $\varphi$.
The effects of these higher dimensional operators are negligible in the vicinity of the GFP.
This can be seen from the change of variable,
\begin{equation}
\label{change_of_variable}
 \varphi \to \left( 2L^4\lambda(\hat{L}) \right)^{-1/4} \varphi \equiv S^{-1/4} \varphi,
\end{equation}
such that operators with dimension greater than four are suppressed by powers of $\hat{L}$.  
This change of variable, Eq.~(\ref{change_of_variable}), also allows us to write the partition function as \cite{Brezin:1985xx},
%
\begin{eqnarray}
\label{partition_function_scaling_variable}
 Z=\mathcal{N} \Omega_{N-1} S^{-N/4} \int_0^{\infty} d\varphi \, \varphi^{N-1} 
 \exp \left(-\frac{1}{2}z\varphi^2 - \varphi^4 \right)
 \equiv \mathcal{N} \Omega_{N-1} S^{-N/4} \bar{\varphi}_{N-1} (z) , 
\end{eqnarray}
where $z = \sqrt{2}\hat{L}^2\hat{m}^2(\hat{L}) \lambda(\hat{L})^{-1/2}$ can be identified with the scaling 
variable.  The determinants in Eq.~(\ref{effective_action_first_expansion_to_mean_field}) will 
renormalise this scaling variable, resulting in logarithmic corrections as detailed below.

We first notice that the integrals in Eq.~(\ref{partition_function_scaling_variable}) can be evaluated, leading to
\begin{eqnarray}
 \bar{\varphi}_0 &=& \frac{\pi}{8} \exp \left( \frac{z^2}{32} \right) \sqrt{|z|} 
 \left[ I_{-1/4}\left( \frac{z^2}{32} \right) - \mathrm{Sgn}(z) \, I_{1/4}\left( \frac{z^2}{32} \right)  \right],\nonumber \\
 \bar{\varphi}_1 &=& \frac{\sqrt{\pi}}{8} \exp \left( \frac{z^2}{16} \right) 
 \left[ 1-\mathrm{Sgn}(z) \, \mathrm{Erf} \left( \frac{|z|}{4} \right) \right], \mbox{ } 
 \label{recursion_formula}
 \bar{\varphi}_{n+2} = -2 \frac{\mathrm{d}}{\mathrm{d} z} \bar{\varphi}_n, 
\end{eqnarray}
where $I_{\nu}$ stands for the modified Bessel function of the first kind.
The leading-order logarithmic corrections to these scaling formulae can be obtained from the one-loop RG equations (RGE's)
\begin{eqnarray}
 -\rho \frac{\mathrm{d}}{\mathrm{d}\rho}Y(\rho) &=& \beta_{YY^2} Y(\rho)^2, 
 \text{  }
 -\rho \frac{\mathrm{d}}{\mathrm{d}\rho}\varphi(\rho) = 2\delta_{Y}Y(\rho)\varphi(\rho),
  \nonumber \\
 -\rho \frac{\mathrm{d}}{\mathrm{d}\rho}\lambda(\rho) &=& \beta_{\lambda \lambda^2}\lambda(\rho)^2 + 
 \beta_{\lambda \lambda Y}\lambda(\rho)Y(\rho) + \beta_{\lambda Y^2} Y(\rho)^2,  \nonumber \\
\label{mass_square_RGE}
  -\rho \frac{\mathrm{d}}{\mathrm{d}\rho}m^2(\rho) &=& 2\left[ \gamma_{Y}Y(\rho) + \gamma_{\lambda} \lambda(\rho) \right]m^2(\rho),
\end{eqnarray}
where the $\beta$'s and the $\gamma$'s are the one-loop RGE coefficients.  They can be calculated straightforwardly.
The solutions to these equations give
\begin{eqnarray}
\label{full_scaling_variable}
 z &=& \left( \frac{4\beta_{\lambda \lambda^2}}{Y(\hat{l})} \right)^{1/2} \left[ Y(\hat{l})(\beta_{+}-\beta_{-}) 
 \right]^{\frac{2\gamma_{\lambda}}{\beta_{\lambda \lambda^2}}} \hat{L}^2
 \left( \hat{m}_b^2 -\hat{m}_c^2 +\frac{A}{\hat{L}^2} \right)\nonumber \\
 && \hspace{0.5 mm}\times \left[ \frac{Y(\hat{l})}{Y(\hat{L})} 
 \right]^{\frac{1}{2}-\frac{2\gamma_Y}{\beta_{YY^2}}-\frac{\beta_{-}\gamma_{\lambda}}{\beta_{YY^2}\beta_{\lambda\lambda^2}}}
 \frac{ \left\{ B_{+} -B_{-}\left[ \frac{Y(\hat{l})}{Y(\hat{L})} \right]^{\frac{\beta_{+}-\beta_{-}}{2\beta_{YY^2}}}  \right\}
 ^{\frac{1}{2}-\frac{2\gamma_{\lambda}}{\beta_{\lambda \lambda^2}}} 
 }{
 \left\{ \beta_{-}B_{+} -\beta_{+}B_{-}\left[ \frac{Y(\hat{l})}{Y(\hat{L})} \right]^{\frac{\beta_{+}-\beta_{-}}{2\beta_{YY^2}}} 
 \right\}^{\frac{1}{2}}},
\end{eqnarray}
where $\beta_{\pm} = (\beta_{YY^2}-\beta_{\lambda \lambda Y}) \pm 
 \sqrt{(\beta_{YY^2}-\beta_{\lambda \lambda Y})^2 - 4 \beta_{\lambda\lambda^2}\beta_{\lambda Y^2}}$
and $B_{\pm} = Y(\hat{l})\beta_{\pm} - 2 \lambda(\hat{l})\beta_{\lambda\lambda^2}$.
This scaling variable contains four free parameters. Two of them, $Y(\hat{l})$ and $\lambda(\hat{l})$,
are integration constants from Eq.~(\ref{mass_square_RGE}).  
One also has to determine the additive renormalisation, $\hat{m}_c^2$, and the coefficient $A$ in the
volume-dependent shift of $\hat{m}^2_c$.  The logarithmic volume dependence is in the renormalised coupling $Y(\hat{L})$.

To test our analytical formulae, 
we perform lattice simulations of a HY model with scalar $O(4)$ symmetry and two fermion flavours 
at weak couplings, 
and confront the logarithmic scaling formula for Binder's cumulant with numerical results.
The continuum action of this model is
\begin{eqnarray}
\label{cont_action}
 S^{cont}[\Phi,\bar{\psi},\psi] &=&\int \mathrm{d}^{4}x \left\{ \frac{1}{2} \left( \partial_{\mu}\Phi\right)^{\dagger}  
 \left( \partial_{\mu}\Phi  \right)
 + \frac{1}{2} m_b^2  \Phi^{\dagger} \Phi 
 +\lambda_b \left( \Phi^{\dagger} \Phi \right)^2 \right\} \nonumber \\
 &&+\int \mathrm{d}^{4}x \left\{ \bar{\Psi} \partial \hspace*{-2.2 mm} \slash \Psi 
 + y_b \left( \bar{\Psi}_{L} \Phi b_{R} + \bar{\Psi}_{L} \tilde{\Phi} t_{R} + h.c. \right) \right\}, 
\end{eqnarray}
where $\Phi^{T} = (\phi_2+i\phi_1,\, \phi_0 - i\phi_3) , \; \Psi^{T} = (t,\,b)$ and $\tilde{\Phi}=i\tau_2\Phi$,
with $\tau_2$ being the second Pauli matrix.  
The bare couplings in this work are always fixed to $y_b = 175/246$, 
which is the ratio between the top-quark and the Higgs vacuum expectation value,
and $\lambda_b = 0.15$.  
We then vary the bare scalar mass, $\hat{m}_b^2$, to scan for the phase structure.
The polynomial Hybrid Monte Carlo algorithm with dynamical overlap
fermions \cite{Frezzotti:1998yp,Gerhold:2010wy} is used to generate scalar field configurations, 
with $7$ choices of the lattice size, $\hat{L}=8,10,12,14,16,20,24$.

\begin{figure}[!t]
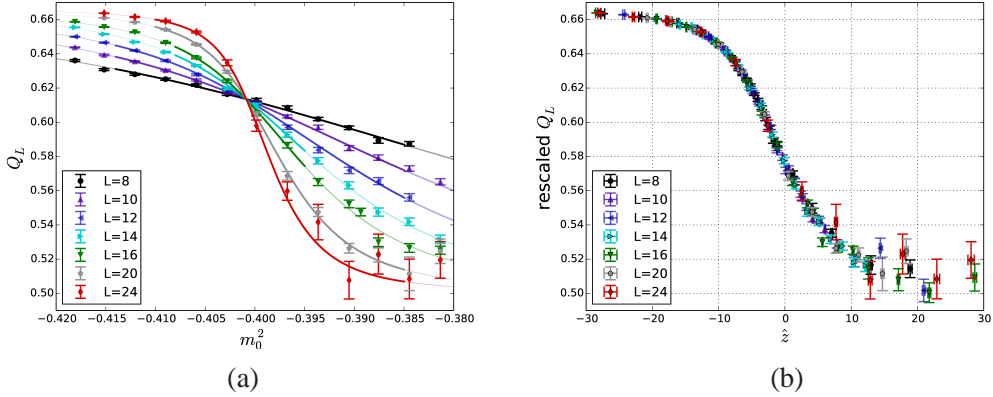

\begin{center}
\includegraphics[width=0.45\columnwidth]{binder_cumulant_niceplot.epsi}
\hspace*{0.5 mm}
\includegraphics[width=0.45\columnwidth]{binder_cumulant_rescale.epsi}
\\
\hspace*{0.5 mm}(a) \hspace*{6.7 cm}(b)
\caption{
(a) Fit to data at all volumes ($\hat{L}=8, \, 10 , \, 12 , \,14, \, 16, \, 20, \, 24$) 
using our analytical formula for Binder's cumulant.  
This plot shows the best fit with $p$ value $\sim 0.5$ and $\chi^2/d.o.f. \sim 1$.
(b) Rescaled plot of Binder's Cumulant using the fit parameters we found in (a).  
  \label{fig:analytic_FV_Binder_Q} 
}
\end{center}
\end{figure}

Binder's cumulant, $Q_L$, can be expressed in terms of the integrals in
Eq.~(\ref{recursion_formula}),
\begin{eqnarray}
\label{Binders_Q}
 Q_L \equiv 1-\frac{\langle \varphi^4 \rangle}{3\langle \varphi^2 \rangle^2}
 = 1-\frac{\bar{\varphi}_7 (z) \, \bar{\varphi}_3 (z)}{3\bar{\varphi}_5(z)^2}.
\end{eqnarray}
We fit the lattice result for $Q_L$ with our analytical formula, and find good agreement, with the 
optimal $p$ value $\sim 0.5$, and $\chi^2/d.o.f. \sim 1$.  Figure~\ref{fig:analytic_FV_Binder_Q}(a)
displays our data points and the curves from the fit.
The free parameters we extract are,
$\hat{m}_c^2 = -0.4008(1)$, $Y(\hat{l}) = 0.42(23)$, $\lambda(\hat{l}) = 0.103(12)$, $A = -1.44(2)$.
Using these results, we can reconstruct the 
scaling variable, $z$. Figure \ref{fig:analytic_FV_Binder_Q}(b)
shows that the data points after rescaling lie on a single curve when plotted against $z$.

\section{Finite Temperature Study with a $\lambda_6 \left( \Phi^{\dagger}\Phi \right)^3$ Term}
\begin{figure}[!t]
\begin{center}
\includegraphics[width=0.45\columnwidth]
{phi6_phase_structure.epsi}
\hspace{0.5 mm}
\includegraphics[width=0.5\columnwidth]
{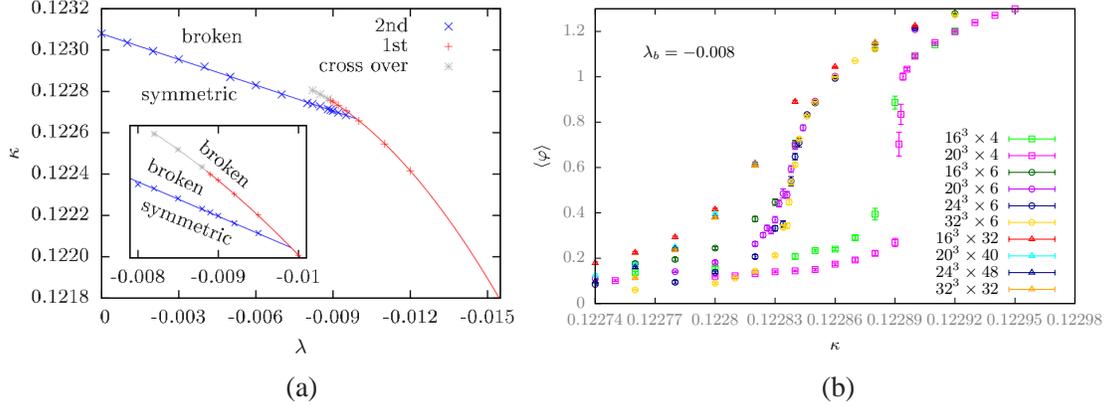}
\\
\hspace*{0.5 cm}(a) \hspace*{6.6 cm}(b)
\\
\caption{
(a) The zero-temperature phase structure of the Higgs-Yukawa model with fixed bare $\lambda_6=0.001$.
The symbol $\lambda$ in this plot is the bare quartic coupling, $\lambda_b$ \cite{Chu:2015nha}.
(b) The plot of the magnetisation with $\lambda_b = -0.008$, $\hat{L}_t=4,6$, anti-periodic boundary condition in time for the fermions,
and the corresponding
zero-temperature results ($\hat{L}_t=32,40,48$) with periodic boundary condition in time for fermions.
  \label{fig:zeroT_phase_structure_and_L20_2nd_and_1st} 
}
\end{center}
\end{figure}

In this section we study the property of the Higgs-Yukawa model with a 
$\lambda_6\int \mathrm{d}^4x \left( \Phi^{\dagger}\Phi \right)^3$
term.  
We use overlap fermions with anti-periodic boundary condition in the temporal direction.
With the zero-temperature phase structure, as shown in Fig.~\ref{fig:zeroT_phase_structure_and_L20_2nd_and_1st}(a),
established in our previous work
\cite{Chu:2015nha}, we proceed to investigate the finite temperature aspect of the model. 
The simulations are performed with fixed bare $\lambda_6=0.001$, and $y_b=175/246$.
We keep $\lambda_b=-0.006$, $-0.007$, $-0.008$, $-0.0085$, $-0.009$ fixed in our simulations
and scan in $\kappa$, defined through $\hat{m}_b^2 = \frac{1-8\kappa^2\lambda_b-8\kappa}{\kappa}$,
to probe the phase structure.
We have studied lattices of the temporal extent $\hat{L}_t=4,6$, and the spatial sizes $\hat{L}=12,16,20,24,32$,
using the magnetisation,
$\langle \varphi \rangle \equiv\frac{1}{\hat{L}^3\hat{L}_t} \langle |\sum_x a\Phi_x| \rangle$,
as the order parameter.

\begin{figure}[!t]
\begin{center}
 \includegraphics[width=0.45\columnwidth]
{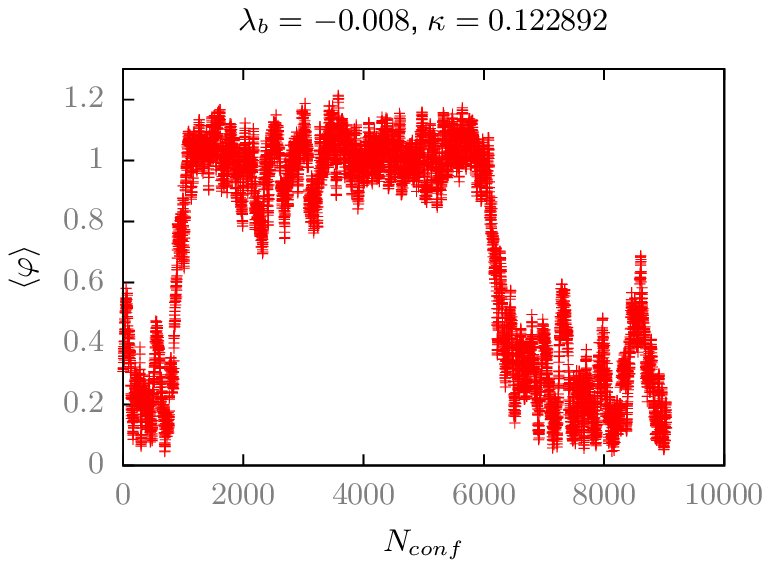}
\hspace{3 mm}
\includegraphics[width=0.45\columnwidth]
{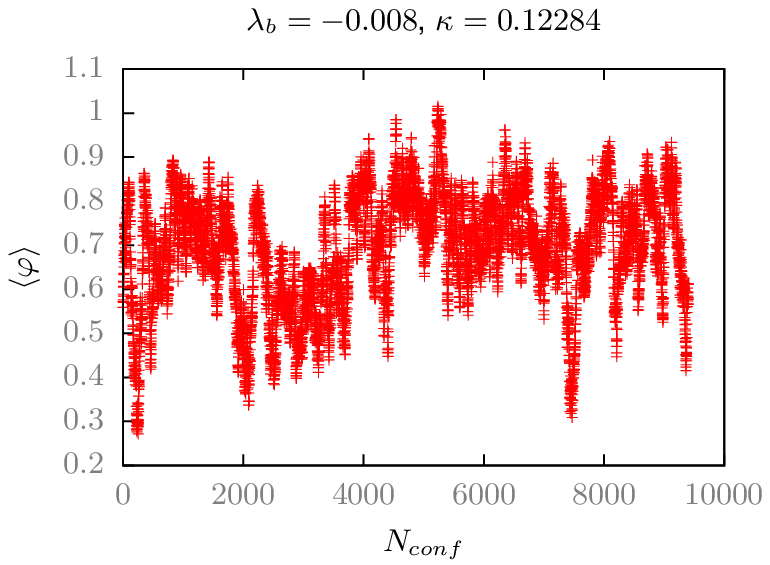}
\\
\hspace*{0.8 cm}(a) \hspace*{6.8 cm}(b)
\\
\caption{The Monte-Carlo time evolution of $\langle \varphi \rangle$ near phase transitions at 
(a) $\kappa= 0.122892$ and the lattice volume $20^3 \times 4$, (b) $\kappa=0.12284$ and the lattice volume $20^3 \times 6$.
\label{fig:temperature_induced_transitions}
}
\end{center}
\end{figure}

Our preliminary result shows evidence that for $\lambda_b=-0.008,-0.0085$, where zero-temperature transitions
are second-order, there can exist temperature-induced first-order phase transitions. 
For example, in Fig.~\ref{fig:zeroT_phase_structure_and_L20_2nd_and_1st}(b) we show when
$\lambda_b = -0.008$, the phase transition is second-order at zero-temperature.  
The transition becomes first-order when $\hat{L}_t = 4$.
As demonstrated in Fig.~\ref{fig:temperature_induced_transitions}(a), the HMC history of the magnetisation exhibits  
coexistence of two states.
Note that this result is preliminary and further detailed scaling tests are needed to confirm this scenario.
We noticed that there are no such coexistence of two states when $\hat{L}_t=6$ as 
shown in Fig.~\ref{fig:temperature_induced_transitions}(b).
We have also performed pure scalar lattice simulations with $\lambda_6 \left( \Phi^{\dagger}\Phi \right)^3$,
where temperature-induced first-order phase transition is not observed when zero-temperature
transitions are second-order.

\section{Summary and Outlook}

We derive FSS formulae including the leading-order logarithmic corrections to the mean-field results for the HY models
near the GFP.  The scaling behaviour of Binder's cumulant is tested against our lattice data at weak couplings, 
and good agreement is found.  In the future, we will perform the same test for other quantities, such as the 
magnetisation and the susceptibility.  Eventually these formulae will be used to explore the phase structure at
strong bare couplings.

As a seperate project, we investigate finite-temperature properties of the HY model in the presence of the 
dimension-6 operator, $\lambda_6 \left( \Phi^{\dagger}\Phi \right)^3$.  In this model, we find preliminary
evidence for first-order temperature-induced phase transitions at the bare parameters where zero-temperature
phase transitions are second-order.  
It will be interesting to investigate this scenario further in the future and see whether it can be established.  
\\ \\
{\bf \large Acknowledgements}
We thank Chung-Wen Kao for valuable discussions.  
The simulations have been performed at the PAX cluster at DESY-Zeuthen
and HPC facilities at National Chiao-Tung University.  
The works are supported by Taiwanese MOST via grant 102-2112-M-009-002-MY3, the support from the 
DAAD-MOST exchange programme via project number 57054177, 
and the support from DFG through the DFG-project  Mu932/4-4.


\bibliographystyle{apsrev} 
\bibliography{refs} 

\end{document}